\documentclass[a4paper,fleqn,usenatbib]{mnras}
\usepackage{graphicx}
\usepackage{txfonts}
\usepackage{morefloats}
\usepackage{float}
\usepackage{color} 
\usepackage{ulem}
\usepackage{gensymb}
\usepackage{times}
\usepackage{epstopdf}
\usepackage[figuresright]{rotating}
\usepackage{longtable}
\usepackage{xspace}

\definecolor{pink}{rgb}{.9,.2,.5}  
\definecolor{purple}{rgb}{.5,.6,.7}

\title[Nebular parameter relations based  on CALIFA galaxies]
      {Effective temperature of  ionizing stars of extragalactic H\,{\sc ii} regions--II:
      nebular parameter relations
      based on CALIFA data}

\author[I.~A.~Zinchenko et al.]
       {I.~A.~Zinchenko$^{1,2}$,  
        O.~L.~Dors$^{3}$,
        G.~F.~H\"{a}gele$^{4,5}$,
        M.~V.~Cardaci$^{4,5}$,
	A.~C. Krabbe$^{3}$\\ 
       $^{1}$ Main Astronomical Observatory
             of National Academy of Sciences of Ukraine,
             27 Zabolotnogo str., 03680 Kiev, Ukraine \\
       $^{2}$ Astronomisches Rechen-Institut, Zentrum f\"{u}r Astronomie 
             der Universit\"{a}t Heidelberg, 
             M\"{o}nchhofstr.\ 12--14, 69120 Heidelberg, Germany \\
       $^{3}$ Universidade do Vale do Paraiba, Av. Shishima Hifumi, 291, Cep12244-000, Sao Jose dos Campos, SP, Brazil \\    
       $^{4}$ Instituto de Astrof\'{i}sica de La Plata (CONICET-UNLP), Argentina \\
       $^{5}$ Facultad de Ciencias Astron\'{o}micas y Geof\'{i}sicas, Universidad Nacional de La Plata, Paseo del Bosque s/n, 1900 La Plata, Argentina \\
}

\date{Accepted 2018 Month 00. Received 2018 August 00; in original form 2018 August 00}

\begin{document}

\maketitle

\begin{abstract}
We calculate the effective temperature ($T_{\rm eff}$) of  ionizing star(s), oxygen abundance 
of the gas phase $(\rm O/H)$, and the ionization parameter $U$ for a sample of  H\,{\sc ii} regions located 
in the disks of 59 spiral galaxies in the $0.005 \: < \: z \: < \: 0.03$ redshift range.  
We use spectroscopic data taken from the CALIFA data release 3 (DR3) and theoretical 
(for $T_{\rm eff}$ and $U$) and empirical (for O/H) calibrations based on strong emission-lines. 
We consider spatial distribution and radial gradients of those parameters in each galactic disk for the objects in our sample. Most of the galaxies in our sample ($\sim70$ \%) shows positive $T_{\rm eff}$ 
radial gradients even though some them exhibit negative or flat ones. 
The median value of the $T_{\rm eff}$ radial gradient is 0.762 kK/$R_{25}$.
We find that radial gradients of both $\log U$ and $T_{\rm eff}$ depend on the oxygen abundance gradient, 
in the sense that the gradient of $\log U$ increases as $\log(\rm O/H)$ gradient increases while there is an anti-correlation between the gradient of $T_{\rm eff}$ and the oxygen abundance gradient. 
Moreover, galaxies with flat oxygen abundance gradients tend to have flat $\log U$ and $T_{\rm eff}$ gradients as well.
Although our results are in agreement with the idea of the existence of positive $T_{\rm eff}$ gradients along the disk of the majority of spiral galaxies, this seems not to be an universal property for these objects.
\end{abstract}


\section{Introduction}

The determination of the effective temperature ($T_{\rm eff}$) of the
ionizing star(s) belonging to H\,{\sc ii} regions is crucial to understand the 
processes that restrict the formation and evolution of massive stars, the physics of stellar atmosphere, 
the excitation of the Interstellar Medium (ISM) as well as the galaxy in which they reside.

For ionizing stars of nearby H\,{\sc ii} regions, located
in the Milky Way and the Magellanic Clouds, the effective temperature 
can be directly estimated by using  their photometric and spectrometric data 
(e.g.\ \citealt{Massey2005, Massey2009, Corti2007, Sota2011, Morrell2014, 
Walborn2014, Lamb2016, Evans2015, MohrSmith2017, Martins2017, Markova2018}). 
However, for the majority of the distant ionizing massive stars,  
$T_{\rm eff}$ can only be  indirectly estimated, e.g.\ from the analysis of 
emission-lines emitted by the nebulae ionized by these stars.
By using this methodology,  proposed by \citet{Zanstra1929}, it is possible to estimate
$T_{\rm eff}$ and its behaviour along the disk of  spiral galaxies 
(see e.g. \citealt{Dors2017} and references therein).

\begin{figure}
\begin{center}
\includegraphics[angle=0,width=1\columnwidth]{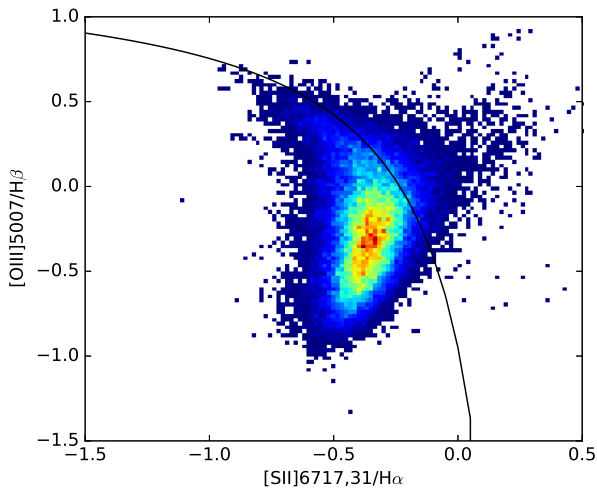}
\end{center}
\caption{BPT diagnostic diagram \citep{BPT} for all the spaxels in our sample. 
The solid line is the boundary between AGN-like and H\,{\sc ii}-like regions defined by 
\citet{Kewley2001}.}
\label{figure:BPT}
\end{figure}

Due to effects of opacity and/or line-blanketing in the stellar atmospheres \citep{Abbott1985, 
Schaerer1994, Martins2005}, stars with higher metallicity ($Z$) trend to present lower values 
of $T_{\rm eff}$ than their counterparts with the same mass but lower $Z$ (e.g.\ \citealt{Mokiem2004, Martins2004}).
It is well known that spiral galaxies exhibit metallicity gradients, in the sense that $Z$ 
decreases with the increment of the galactocentric radius (e.g.\ \citealt{Pilyugin2004}).
Therefore, assuming that stars are formed with 
an universal stellar upper mass limit of the Initial Mass Function (e.g.\ \citealt{Bastian2010}), it is expected a positive gradient
of $T_{\rm eff}$ in the disks of spiral galaxies. In fact, \citet{Shields1978} interpreted that the enhancement of the equivalent width of the H$\beta$ emission-line with the galactocentric distance for a sample of H\,{\sc ii} regions in M\,101 could be due to a positive $T_{\rm eff}$ gradient 
(see also \citealt{Vilchez1988, Henry1995, Dors2003, Dors2005}). In spite of  these gradients should exist in most of the spiral galaxies, they were not found in early studies (e.g.\ \citealt{Fierro1986, Evans1986}).
Recently, \citet{Dors2017} studied the $T_{\rm eff}$ variation as a function of
the galactocentric distance for H\,{\sc ii} regions belonging to 14 spiral galaxies using a
new theoretical calibration between the observed emission-line ratio 
$R$=log([O\,{\sc ii}]($\lambda$$\lambda$3726+29)/[O\,{\sc iii}]$\lambda$5007) and $T_{\rm eff}$ \cite[relation proposed by][]{Dors2003}. These authors found positive 
gradients for 11 of these galaxies, null gradients for two and a negative gradient for the other one (see also \citealt{PerezMontero2009}).   
In particular, the first 
negative $T_{\rm eff}$ gradient was found  for the Milky Way by \citet{Morisset2004}.  Additional analysis taking into account
a larger number of galaxies is necessary to ascertain if the $T_{\rm eff}$ gradient is an universal property of spiral galaxies.
  
The knowledge of the relation between different physical parameters  is essential 
to comprehend which mechanisms drive the  formation and evolution of galaxies. 
For example,  in the seminal paper, \citet{Lequeux1979} calculated the metallicity (traced by the
ratio between oxygen and hydrogen abundances) and the total galaxy mass ($M_{\rm T}$; obtained from atomic hydrogen velocity maps) 
for eight irregular and blue compact galaxies and found a clear relation between these 
parameters (see also \citealt{Kinman1981, Peimbert1982, Rubin1984, Skillman1992, Tremonti2004, Pilyugin2004,
Sanchez2013}, among others). 
Thereafter, \citet{Ellison2008} showed that the $M_{\rm T}$-$Z$ relation is affected by a dependence
of the metallicity on the star formation rate (SFR), establishing the $M_{\rm T}$-$Z$-SFR relation.
This result was confirmed by \citet{LaraLopez2010,Mannucci2010}. However, \citet{Sanchez2017} in their recent study,
based on integral field spectroscopy data, did not find any significant dependence of the $M_{\rm T}$-$Z$ 
relation with the SFR, but they did not exclude the existence of such relation 
on local scales, e.g.\ in the central regions of the galaxies.
Therefore, the existence of the universality of $T_{\rm eff}$ gradients, together with the $M_{\rm T}$-$Z$-SFR relation,  
would produce additional and fundamental concepts of the physical processes taking place in galaxies,  
important insights into how the formation and evolution
of massive stars occur as well as their interaction with the ISM.

In this work, we use the methodology presented by \cite[][hereafter Paper I]{Dors2017}, 
 to estimate $T_{\rm eff}$ of extragalactic
H\,{\sc ii} regions located in a large sample of spiral galaxies. We have taken advantage of the existence of an homogeneous sample of spectroscopic data of H\,{\sc ii} regions obtained as part of the Calar Alto Legacy Integral Field Area Survey\footnote{www.http://califa.caha.es/} 
(CALIFA, \citealt{Sanchez2012}), which is ideal to investigate global scaling relations between galaxy properties (see e.g.\ \citealt{Ellison2018}).
The main goals of the present study are to investigate if $T_{\rm eff}$ gradients are universal properties of spiral galaxies, and
the existence of any correlation between $T_{\rm eff}$ and nebular parameters such as the ionization parameter or the oxygen abundance.
This paper is organized as follows. The methodology assumed to calculate $T_{\rm eff}$ and the observational
data used  along this work are described in Sec.~\ref{met}.  In Sect.~\ref{res-disc} the results and  discussion of the outcome are presented. 
Finally, conclusions are given in Sect.~\ref{conc}.

\section{Methodology}
\label{met}

\subsection{Sample}

We used publicly available spectra from the integral field spectroscopic CALIFA
survey data release 3 \citep[DR3;][]{Sanchez2016,Sanchez2012,CALIFA2014} based on
observations with the PMAS/PPAK integral field spectrophotometer mounted on the 
Calar Alto 3.5-meter telescope. CALIFA DR3 provides wide-field IFU data for 667 objects in total.  
The data for each galaxy consist of two 
datacubes, which cover the spectral regions of
4300--7000\,$\AA$ at a spectral resolution of $R \sim 850$ (setup V500) and of 3700--5000\,$\AA$ 
at $R \sim 1650$ (setup V1200). For the galaxies with both V500 and V1200 datacubes available, there are COMB datacubes for 446 galaxies which are a combination of V500 and V1200 datacubes 
covering the 3700--7000\,$\AA$ spectral range. In this study we used these COMB datacubes.

\begin{figure}
  \begin{center}
  \includegraphics[width=0.95\linewidth]{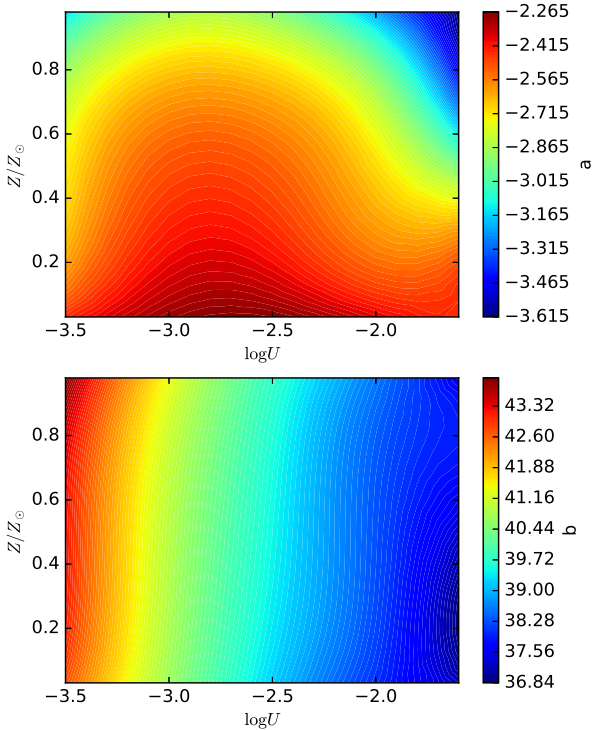}
  \end{center}
\caption{Coefficients $a$ (upper panel) and $b$ (lower panel) of the linear regression $T_{\rm eff} = a \times R+b$ as a function of $\log U$ and $Z/Z_{\odot}$.
}
\label{figure:coefs}
\end{figure}

The sample of galaxies is described in detail in \citet[][in prep.]{Zinchenko2018}. Briefly,
we selected isolated galaxies with inclination less than $60\degree$. Galaxies with 
insufficient number of spaxels with measured oxygen abundance were excluded from our sample.
We also rejected from our sample galaxies with oxygen abundance measurements for less than 50 spaxels
and/or galaxies for which spaxels with oxygen abundance measurements cover a range of galactocentric 
distances lower than $\sim$ 1/3 of the its optical radius.
Stellar masses, derived from UV-to-NIR photometry, has been taken from \citet{CALIFA2014}.
Our final sample contains 59 galaxies and 49067 spaxels.

The final spatial resolution of the CALIFA data is set by the fiber size of the PMAS/PPAK integral 
field spectrophotometer and it is of the order of 3\,arcsec \citep{Husemann2013}.
Thus, the spectrum of each spaxel corresponds to the spectrum emitted by a region  
with a diameter varying from $\sim$300 pc to $\sim$1.5 kpc depending on the distance to the galaxy\footnote{We assumed a spatially flat cosmology with
$H_{0}$\,=\,71 $ \rm km\:s^{-1} Mpc^{-1}$, $\Omega_{m}=0.270$, and $\Omega_{\rm vac}=0.730$  \citep{Wright2006}}.
Therefore, each observed spectrum comprises the flux of a complex of H\,{\sc ii} regions and 
the physical properties derived represent an averaged value (see discussion in Paper I).

\begin{figure*}
\begin{center}
\includegraphics[width=0.9\linewidth]{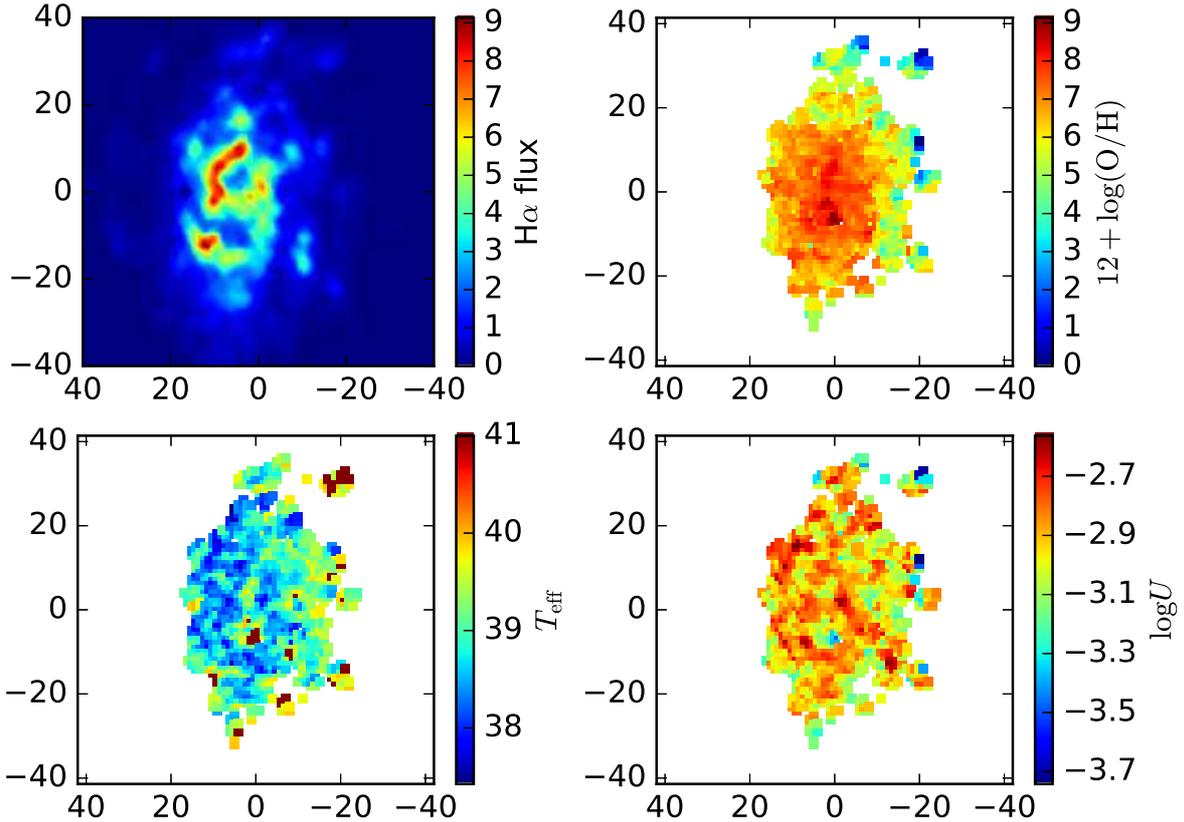}
\end{center}
\caption{Maps of the H$\alpha$ emission-line flux (top-left panel), oxygen abundance (top-right panel),
$T_{\rm eff}$ (bottom-left panel), and $\log U$ (bottom-right panel) for NGC~237.}
\label{figure:NGC237maps}
\end{figure*}

\subsection{The emission line fluxes}
\label{data}

The spectrum of each spaxel from the CALIFA DR3 datacubes is processed in the same way as 
described in \citet{Zinchenko2016}. Briefly, 
the stellar background in all spaxels is fitted using the public version of the STARLIGHT code 
\citep{CidFernandes2005,Mateus2006,Asari2007} adapted for execution in the NorduGrid
ARC\footnote{http://www.nordugrid.org/} environment of the Ukrainian National Grid.
To fit the stellar spectra we used 45 synthetic simple stellar population (SSP) spectra
from the evolutionary synthesis models by \citet{BC03} with ages from 1~Myr up to 13~Gyr 
and metallicities $Z = 0.004$, 0.02, and 0.05. We adopted the reddening law of \citet[]{CCM} with $R_V = 3.1$.
The resulting stellar radiation contribution is subtracted from the observed spectrum 
in order to  measure and analyse the line emission from the gaseous
component. The line intensities were measured using single Gaussian line
profile fittings on the pure emission spectra.

The total [O\,{\sc iii}]$\lambda$$\lambda$4959,5007 flux has been estimated as 
$1.33 \: \times$~[O\,{\sc iii}]$\lambda$5007 instead of as the sum of the fluxes of both lines. 
These lines originate from transitions from the same energy level, so their flux ratio can be 
determined by the transition probability ratio, which is very close to 3 \citep{Storey2000}.
The strongest line, [O\,{\sc iii}]$\lambda$5007, can be measured with higher precision than 
the weakest one. This is particularly important for high-metallicity
H\,{\sc ii} regions, which have weak high-excitation emission-lines. 
Similarly, the [N\,{\sc ii}]$\lambda$$\lambda$6548,6584 lines also originate from transitions from the 
same energy level and the transition probability ratio for those lines is again
close to 3 \citep{Storey2000}. Therefore, we estimated its total flux as 
$1.33$~[N\,{\sc ii}]$\lambda$6584. For each spectrum, we measure the fluxes of the 
[O\,{\sc ii}]$\lambda\,\lambda$3727,3729,
H$\beta$,  
[O\,{\sc iii}]$\lambda$5007,
H$\alpha$,  
[N\,{\sc ii}]$\lambda$6584, and
[S\,{\sc ii}]$\lambda$6717, 6731.
For the further analysis we selected only those spectra, for which the signal-to-noise ratio is larger than $5$ for each emission-line listed above.
The measured line fluxes are corrected for interstellar reddening using the theoretical H$\alpha$ to H$\beta$ ratio  
assuming the standard value of H$\alpha$/H$\beta$ = 2.86, and the analytical approximation of the Whitford interstellar 
reddening law from \citet{Izotov1994}.  When the measured value of H$\alpha$/H$\beta$ is lower than 2.86 the reddening is adopted to be zero.

Following Paper I, we apply the 
$\log$([O\,{\sc iii}]$\lambda$5007/H$\beta$) -- $\log$([S\,{\sc ii}]$\lambda\lambda$6717,6731/H$\alpha$) 
criterion proposed by \cite{Kewley2001} to separate objects for which the main ionization source are
massive stars from those whose main ionization source are shocks of gas and/or active 
galactic nuclei (AGNs).  We consider only the objects located below the separation line defined by \citet{Kewley2001}, 
i.e. 39431 spaxels.
In Fig.~\ref{figure:BPT}, the BPT diagnostic diagram \citep{BPT} for the all the spaxels in our sample is presented.
 
\subsection{Nebular parameter determinations}
\label{method}


In order to estimate $T_{\rm eff}$, we adopted the same method proposed in Paper I, where 
a new calibration between $T_{\rm eff}$ and the $R$ = $\log$([O\,{\sc ii}]$\lambda\lambda$3727,3729/[O\,{\sc iii}]$\lambda$5007)
ratio was proposed. The method consists of three steps:
a) to estimate the metallicity $Z$ of star forming regions,
b) to estimate $U$ from $Z$ and [S\,{\sc ii}]$\lambda\lambda$6717,6731
and H$\alpha$ emission lines, and 
c) to estimate $T_{\rm eff}$ from $U$, $Z$, and $R$. 

Entering into details, the first step consists in calculating the metallicity of the gas  traced by the oxygen abundance 
in relation to the hydrogen one, in units of  12+$\log(\rm O/H)$. It is carried out using the R$_{3D}$ empirical calibration given 
by \citet{PilyuginGrebel2016}. These authors derived oxygen abundances based on direct estimations of the electron temperatures 
for a large sample of H\,{\sc ii} regions
and they obtained relations between these abundances and  the emission line flux ratios of oxygen and nitrogen in relation to H$\beta$ defined as:
$R_2 = $[O\,{\sc ii}]$\lambda\lambda$3727,3729/H$\beta$,  
$R_3 = $[O\,{\sc iii}]$\lambda\lambda$4959,5007/H$\beta$,
$N_2 = $[N\,{\sc ii}]$\lambda\lambda$6548,6584/H$\beta$. 
To use this calibration it is necessary to define which branch of the curve must be considered,  due to
the degeneracy in the calibrations.
For H\,{\sc ii} regions with $\log N_{2} \ge -0.6$ the upper branch is assumed and the relation is the following:
\begin{eqnarray}
       \begin{array}{lll}
     {\rm (O/H)}_{R,U}  & = &   8.589 + 0.022 \, \log (R_{3}/R_{2}) + 0.399 \, \log N_{2}   \\  
                          & + &  (-0.137 + 0.164 \, \log (R_{3}/R_{2}) + 0.589 \log N_{2})   \\ 
                          & \times &  \log R_{2},   \\ 
     \end{array}
\label{equation:r3du}
\end{eqnarray}
where (O/H)$_{R,U}$ means 12 +log(O/H)$_{R,U}$. 

For H\,{\sc ii} regions with $\log N_{2} < -0.6$, the lower branch is assumed and 
the relation is
\begin{eqnarray}
       \begin{array}{lll}
     {\rm (O/H)}_{R,L}  & = &   7.932 + 0.944 \, \log (R_{3}/R_{2}) + 0.695 \, \log N_{2}   \\  
                          & + &  (0.970 - 0.291 \, \log (R_{3}/R_{2}) - 0.019 \log N_{2})   \\ 
                          & \times & \log R_{2},   \\ 
     \end{array}
\label{equation:r3dl}
\end{eqnarray}
where (O/H)$_{R,L}$ means 12 +log(O/H)$_{R,L}$. To convert the oxygen abundance to  metallicity $Z/Z_{\odot}$, 
we assumed the solar oxygen abundance  12+$\log(\rm O/H)_{\odot} = 8.69$ \citep{AllendePrieto2001}.

Following Paper I, the logarithm of the ionization parameter, $\log U$, is calculated as:
\begin{equation}
\log U = c \times S2 + d,
\label{equation:Teff}
\end{equation} 
where $S2$ = $\log$([S\,{\sc ii}]$\lambda\lambda$6717,6731/H$\alpha$), 
$c = -0.26 \times (Z/Z_{\odot}) - 1.54$, and 
$d = -3.69 \times (Z/Z_{\odot})^2 + 5.11 \times (Z/Z_{\odot}) - 5.26$.

The  $T_{\rm eff}$-$R$  relation  derived in Paper I is
\begin{equation}
\label{eq1}
T_{\rm eff} = a \times R + b,
\end{equation} 
where $R$ = $\log$([O\,{\sc ii}]$\lambda\lambda$3727,3729/[O\,{\sc iii}]$\lambda$5007).
Fig.~\ref{figure:coefs} shows the values of the $a$ and $b$ coefficients as a function of  
$Z/Z_{\odot}$ and $\log U$. In this figure, the values of the coefficients are calculated interpolating the relations given in Table~2 of Paper I.

It is worth to mention that the expected maximum effective temperature of a young stellar cluster
is $\sim$ 50 kK \citep[e.g.,][]{Martins2005,SimonDiaz2014,Tramper2014,Walborn2014,Wright2015,Crowther2016,Martins2017,Holgado2018}
Meanwhile,
the $T_{\rm eff}-R$ relation can be applied only for $T_{\rm eff} \lid 40$~kK because,
for higher $T_{\rm eff}$ values, small variations of $R$ produce extremely large uncertainties in $T_{\rm eff}$
estimations  \citep{Dors2017}. \citet{Kennicutt2000} derived a calibration between $T_{\rm eff}$ and the 
He\,{\sc ii}$\lambda$5876/H$\beta$ and He\,{\sc ii}$\lambda$6678/H$\beta$ emission-line ratios,  and they also pointed out  a similar diffi\-cul\-ty in de\-ri\-ving
effective temperature values  higher than $\sim$40kK. Thus, in this study we consider only objects for which we derive a  $T_{\rm eff}$ value lower or equal to  40~kK. 
It should also be noted that $T_{\rm eff}$ values higher than 40~kK were derived 
only for about 15\% of the  H\,{\sc ii} regions in our sample.
Fig.~\ref{figure:NGC237maps} shows an example of the obtained maps for the H$\alpha$ 
emission line flux, oxygen abundance, $T_{\rm eff}$, and $\log U$ for one of the galaxies in our sample (NGC~237).

\subsection{Radial gradients}

 
Using the methodology and the observational data presented above,
we calculate radial gradients for $T_{\rm eff}$, $\log U$ and  $12+\log({\rm O/H})$  
along the disk of the galaxies in our sample.
For each galaxy, we fitted the radial distributions of these parameters 
by the use of the following relation: 
\begin{equation}
\label{eq2}
Y  = Y_0 + grad\,Y \times R/R_{25} ,
\end{equation} 
where $Y$ is a given parameter, $Y_0$ is the extrapolated value of this parameter to the galactic center,  
$grad\,Y$ is the slope of the distribution expressed in Y units per  optical radius $R_{25}$.
The  radial gradients were estimated using the data  in 
galactocentric distances $0.2 \: R_{25} \:  < \: R \: < \: R_{25}$.

\begin{figure}
\begin{center}
\includegraphics[angle=0,width=0.9\columnwidth]{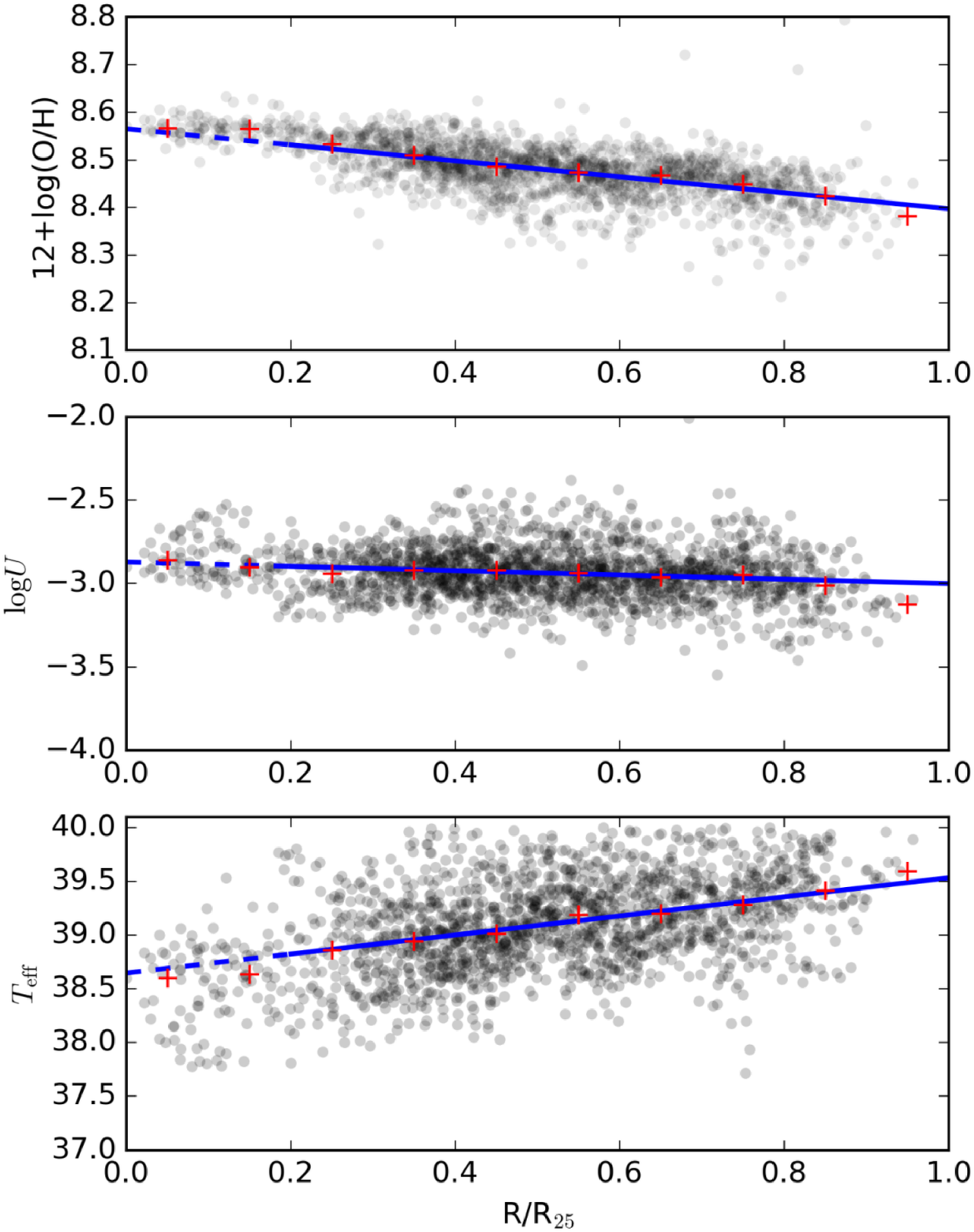}
\end{center}
\caption{Radial gradients of  effective temperature ($T_{\rm eff}$), logarithm of the ionization parameter ($\log U$)
and of oxygen abundance (12+log(O/H) for NGC\,2730, one galaxy of the CALIFA sample. 
The gray filled circles represent the observational data. The blue solid line is the best fit for the data, 
the blue dashed line is the extrapolation to the center. The red plus signs are median values in bins of 0.1 $R_{25}$.}
\label{fig0_oli}
\end{figure}

In Fig.~\ref{fig0_oli}, an example of the radial gradients of $T_{\rm eff}$,  $\log U$, and $12+\log({\rm O/H})$ 
is presented for the spiral galaxy NGC\,2730. This galaxy has a clear positive radial gradient of $T_{\rm eff}$ 
and negative radial gradients of $\log U$ and $12+\log({\rm O/H})$. 
Same plots for the radial gradients together with the $12+\log({\rm O/H})$ vs.\ $\log U$, 
$12+\log({\rm O/H})$ vs.\ $T_{\rm eff}$ diagrams for each galaxy in our sample are available in a supplementary material.

Table~\ref{table:sample} presents our sample of galaxies and the best fit of the radial distribution of 
$T_{\rm eff}$, $\log U$ and  $12+\log({\rm O/H})$ for each galaxy.

\setcounter{table}{0}
\begin{table*}
\caption[]{\label{table:sample}
List of the selected galaxies from the CALIFA survey 
}
\begin{center}
\begin{tabular}{lccccccccc} \hline \hline
Name                       &
$12+\log({\rm O/H})_0$     &  
grad $\log({\rm O/H})$     &
$\log U$                   &
grad $\log U$              &
$T_{\rm eff}$              &
grad $T_{\rm eff}$         &
\\  
                           &
dex                        &
dex/$R_{25}$               &
dex                        &
dex/$R_{25}$               &
kK                         &
kK/$R_{25}$  \\
\hline 
NGC 1 &   8.615 $\pm$ 0.006 &  -0.029 $\pm$ 0.012 &  -2.865 $\pm$ 0.026 &  0.073 $\pm$ 0.052 &  38.802 $\pm$ 0.120 & -0.317 $\pm$ 0.234 \\
NGC 23 &   8.730 $\pm$ 0.010 &  -0.209 $\pm$ 0.024 &  -3.400 $\pm$ 0.058 &  0.751 $\pm$ 0.146 &  39.393 $\pm$ 0.160 & -0.734 $\pm$ 0.369 \\
NGC 180 &   8.705 $\pm$ 0.034 &  -0.217 $\pm$ 0.049 &  -2.629 $\pm$ 0.083 & -0.038 $\pm$ 0.120 &  38.088 $\pm$ 0.337 &  0.380 $\pm$ 0.484 \\
NGC 234 &   8.617 $\pm$ 0.004 &  -0.062 $\pm$ 0.009 &  -2.696 $\pm$ 0.014 & -0.382 $\pm$ 0.032 &  37.938 $\pm$ 0.071 &  1.205 $\pm$ 0.161 \\
NGC 237 &   8.680 $\pm$ 0.003 &  -0.317 $\pm$ 0.007 &  -2.835 $\pm$ 0.011 & -0.165 $\pm$ 0.023 &  38.139 $\pm$ 0.038 &  1.405 $\pm$ 0.077 \\
NGC 257 &   8.696 $\pm$ 0.014 &  -0.255 $\pm$ 0.021 &  -2.806 $\pm$ 0.044 &  0.020 $\pm$ 0.068 &  38.152 $\pm$ 0.160 &  0.585 $\pm$ 0.248 \\
NGC 309 &   8.781 $\pm$ 0.015 &  -0.432 $\pm$ 0.030 &  -2.503 $\pm$ 0.066 & -0.435 $\pm$ 0.133 &  36.933 $\pm$ 0.218 &  2.893 $\pm$ 0.439 \\
NGC 477 &   8.543 $\pm$ 0.010 &  -0.058 $\pm$ 0.014 &  -2.717 $\pm$ 0.025 & -0.201 $\pm$ 0.034 &  38.193 $\pm$ 0.085 &  0.830 $\pm$ 0.116 \\
NGC 776 &   8.647 $\pm$ 0.007 &   0.002 $\pm$ 0.014 &  -2.826 $\pm$ 0.048 &  0.198 $\pm$ 0.101 &  38.149 $\pm$ 0.210 &  0.516 $\pm$ 0.440 \\
NGC 941 &   8.561 $\pm$ 0.006 &  -0.394 $\pm$ 0.015 &  -2.816 $\pm$ 0.018 & -0.410 $\pm$ 0.047 &  38.698 $\pm$ 0.050 &  1.328 $\pm$ 0.134 \\
NGC 991 &   8.532 $\pm$ 0.008 &  -0.269 $\pm$ 0.019 &  -2.951 $\pm$ 0.028 & -0.034 $\pm$ 0.067 &  38.945 $\pm$ 0.067 &  0.515 $\pm$ 0.163 \\
NGC 1070 &   8.626 $\pm$ 0.032 &  -0.035 $\pm$ 0.111 &  -2.673 $\pm$ 0.103 & -0.150 $\pm$ 0.349 &  38.754 $\pm$ 0.492 & -1.394 $\pm$ 1.669 \\
NGC 1094 &   8.656 $\pm$ 0.004 &  -0.117 $\pm$ 0.010 &  -3.127 $\pm$ 0.021 &  0.291 $\pm$ 0.052 &  38.955 $\pm$ 0.070 &  0.570 $\pm$ 0.170 \\
NGC 1659 &   8.621 $\pm$ 0.006 &  -0.184 $\pm$ 0.012 &  -2.777 $\pm$ 0.017 & -0.113 $\pm$ 0.034 &  38.233 $\pm$ 0.053 &  0.978 $\pm$ 0.105 \\
NGC 1667 &   8.657 $\pm$ 0.002 &  -0.067 $\pm$ 0.005 &  -2.847 $\pm$ 0.013 & -0.084 $\pm$ 0.030 &  38.702 $\pm$ 0.049 &  0.059 $\pm$ 0.108 \\
NGC 2347 &   8.689 $\pm$ 0.007 &  -0.279 $\pm$ 0.011 &  -2.860 $\pm$ 0.023 &  0.013 $\pm$ 0.036 &  37.968 $\pm$ 0.069 &  1.601 $\pm$ 0.108 \\
NGC 2487 &   8.639 $\pm$ 0.064 &  -0.054 $\pm$ 0.123 &  -2.706 $\pm$ 0.166 & -0.041 $\pm$ 0.318 &  39.319 $\pm$ 0.992 & -1.832 $\pm$ 1.900 \\
NGC 2530 &   8.557 $\pm$ 0.006 &  -0.251 $\pm$ 0.010 &  -2.754 $\pm$ 0.018 & -0.259 $\pm$ 0.030 &  38.098 $\pm$ 0.046 &  1.788 $\pm$ 0.078 \\
NGC 2540 &   8.626 $\pm$ 0.004 &  -0.154 $\pm$ 0.006 &  -2.904 $\pm$ 0.015 &  0.003 $\pm$ 0.026 &  38.434 $\pm$ 0.054 &  0.715 $\pm$ 0.091 \\
NGC 2604 &   8.524 $\pm$ 0.008 &  -0.401 $\pm$ 0.020 &  -2.834 $\pm$ 0.020 & -0.488 $\pm$ 0.048 &  39.348 $\pm$ 0.042 &  0.586 $\pm$ 0.112 \\
NGC 2730 &   8.565 $\pm$ 0.003 &  -0.167 $\pm$ 0.006 &  -2.870 $\pm$ 0.012 & -0.129 $\pm$ 0.021 &  38.646 $\pm$ 0.030 &  0.886 $\pm$ 0.054 \\
NGC 2906 &   8.643 $\pm$ 0.005 &   0.001 $\pm$ 0.011 &  -2.911 $\pm$ 0.031 &  0.025 $\pm$ 0.063 &  38.494 $\pm$ 0.132 &  0.328 $\pm$ 0.265 \\
NGC 2916 &   8.609 $\pm$ 0.015 &  -0.108 $\pm$ 0.028 &  -2.728 $\pm$ 0.043 & -0.085 $\pm$ 0.076 &  37.681 $\pm$ 0.125 &  1.895 $\pm$ 0.223 \\
NGC 3057 &   8.318 $\pm$ 0.007 &  -0.194 $\pm$ 0.014 &  -3.132 $\pm$ 0.020 & -0.110 $\pm$ 0.039 &  39.899 $\pm$ 0.060 & -0.324 $\pm$ 0.126 \\
NGC 3381 &   8.592 $\pm$ 0.004 &  -0.226 $\pm$ 0.009 &  -2.927 $\pm$ 0.012 & -0.029 $\pm$ 0.027 &  38.495 $\pm$ 0.040 &  1.064 $\pm$ 0.092 \\
NGC 3614 &   8.651 $\pm$ 0.014 &  -0.358 $\pm$ 0.032 &  -2.764 $\pm$ 0.038 & -0.272 $\pm$ 0.087 &  37.984 $\pm$ 0.156 &  1.789 $\pm$ 0.354 \\
NGC 3687 &   8.697 $\pm$ 0.003 &  -0.251 $\pm$ 0.007 &  -3.040 $\pm$ 0.015 &  0.368 $\pm$ 0.037 &  38.918 $\pm$ 0.049 &  0.052 $\pm$ 0.123 \\
NGC 3811 &   8.649 $\pm$ 0.003 &  -0.126 $\pm$ 0.006 &  -2.861 $\pm$ 0.015 &  0.040 $\pm$ 0.026 &  37.983 $\pm$ 0.057 &  1.382 $\pm$ 0.098 \\
NGC 4961 &   8.533 $\pm$ 0.005 &  -0.297 $\pm$ 0.009 &  -2.887 $\pm$ 0.012 & -0.338 $\pm$ 0.022 &  38.814 $\pm$ 0.033 &  1.221 $\pm$ 0.070 \\
NGC 5000 &   8.617 $\pm$ 0.012 &  -0.075 $\pm$ 0.015 &  -2.638 $\pm$ 0.087 & -0.220 $\pm$ 0.102 &  37.199 $\pm$ 0.215 &  1.764 $\pm$ 0.251 \\
NGC 5016 &   8.711 $\pm$ 0.016 &  -0.254 $\pm$ 0.029 &  -2.620 $\pm$ 0.089 & -0.281 $\pm$ 0.161 &  38.128 $\pm$ 0.223 &  1.188 $\pm$ 0.402 \\
NGC 5205 &   8.559 $\pm$ 0.033 &   0.001 $\pm$ 0.065 &  -2.723 $\pm$ 0.084 &  0.146 $\pm$ 0.168 &  37.566 $\pm$ 0.435 &  0.688 $\pm$ 0.863 \\
NGC 5320 &   8.637 $\pm$ 0.003 &  -0.210 $\pm$ 0.006 &  -2.817 $\pm$ 0.010 & -0.121 $\pm$ 0.018 &  38.334 $\pm$ 0.037 &  0.762 $\pm$ 0.067 \\
NGC 5406 &   8.650 $\pm$ 0.009 &  -0.074 $\pm$ 0.017 &  -2.516 $\pm$ 0.046 & -0.458 $\pm$ 0.084 &  37.698 $\pm$ 0.166 &  1.248 $\pm$ 0.305 \\
NGC 5480 &   8.601 $\pm$ 0.003 &  -0.089 $\pm$ 0.009 &  -2.842 $\pm$ 0.015 & -0.172 $\pm$ 0.039 &  38.262 $\pm$ 0.053 &  1.161 $\pm$ 0.135 \\
NGC 5520 &   8.620 $\pm$ 0.002 &  -0.097 $\pm$ 0.004 &  -2.979 $\pm$ 0.009 &  0.126 $\pm$ 0.015 &  38.825 $\pm$ 0.029 &  0.203 $\pm$ 0.044 \\
NGC 5633 &   8.675 $\pm$ 0.003 &  -0.166 $\pm$ 0.006 &  -2.795 $\pm$ 0.013 & -0.235 $\pm$ 0.028 &  38.295 $\pm$ 0.048 &  0.972 $\pm$ 0.101 \\
NGC 5720 &   8.531 $\pm$ 0.030 &   0.021 $\pm$ 0.041 &  -2.856 $\pm$ 0.089 & -0.105 $\pm$ 0.119 &  39.715 $\pm$ 0.313 & -0.353 $\pm$ 0.420 \\
NGC 5732 &   8.605 $\pm$ 0.005 &  -0.180 $\pm$ 0.008 &  -2.840 $\pm$ 0.012 & -0.017 $\pm$ 0.018 &  38.287 $\pm$ 0.048 &  0.944 $\pm$ 0.072 \\
NGC 5957 &   8.688 $\pm$ 0.009 &  -0.180 $\pm$ 0.019 &  -2.830 $\pm$ 0.046 &  0.006 $\pm$ 0.099 &  38.816 $\pm$ 0.164 & -0.403 $\pm$ 0.353 \\
NGC 6004 &   8.639 $\pm$ 0.007 &  -0.039 $\pm$ 0.022 &  -2.647 $\pm$ 0.043 & -0.237 $\pm$ 0.120 &  38.085 $\pm$ 0.173 &  1.160 $\pm$ 0.487 \\
NGC 6063 &   8.561 $\pm$ 0.008 &  -0.065 $\pm$ 0.010 &  -2.940 $\pm$ 0.019 &  0.054 $\pm$ 0.025 &  39.200 $\pm$ 0.063 & -0.175 $\pm$ 0.082 \\
NGC 6154 &   8.619 $\pm$ 0.019 &  -0.019 $\pm$ 0.025 &  -2.806 $\pm$ 0.093 & -0.153 $\pm$ 0.119 &  39.245 $\pm$ 0.322 & -0.314 $\pm$ 0.410 \\
NGC 6155 &   8.592 $\pm$ 0.005 &  -0.088 $\pm$ 0.011 &  -2.917 $\pm$ 0.014 &  0.043 $\pm$ 0.031 &  38.520 $\pm$ 0.056 &  0.265 $\pm$ 0.120 \\
NGC 6301 &   8.622 $\pm$ 0.016 &  -0.086 $\pm$ 0.021 &  -3.102 $\pm$ 0.057 &  0.320 $\pm$ 0.074 &  40.178 $\pm$ 0.355 & -1.421 $\pm$ 0.431 \\
NGC 6497 &   8.652 $\pm$ 0.006 &  -0.008 $\pm$ 0.009 &  -2.785 $\pm$ 0.039 & -0.162 $\pm$ 0.059 &  38.926 $\pm$ 0.144 & -0.102 $\pm$ 0.220 \\
NGC 6941 &   8.581 $\pm$ 0.019 &   0.035 $\pm$ 0.022 &  -3.135 $\pm$ 0.082 &  0.485 $\pm$ 0.098 &  37.614 $\pm$ 0.459 &  0.973 $\pm$ 0.551 \\
NGC 7321 &   8.633 $\pm$ 0.003 &  -0.077 $\pm$ 0.005 &  -3.016 $\pm$ 0.014 &  0.090 $\pm$ 0.023 &  38.660 $\pm$ 0.043 &  0.820 $\pm$ 0.071 \\
NGC 7489 &   8.626 $\pm$ 0.007 &  -0.396 $\pm$ 0.012 &  -2.878 $\pm$ 0.016 & -0.156 $\pm$ 0.026 &  38.525 $\pm$ 0.060 &  1.287 $\pm$ 0.115 \\
NGC 7653 &   8.673 $\pm$ 0.003 &  -0.230 $\pm$ 0.006 &  -2.903 $\pm$ 0.012 &  0.118 $\pm$ 0.026 &  38.444 $\pm$ 0.043 &  0.400 $\pm$ 0.091 \\
NGC 7716 &   8.613 $\pm$ 0.005 &  -0.109 $\pm$ 0.010 &  -2.890 $\pm$ 0.019 &  0.064 $\pm$ 0.039 &  38.621 $\pm$ 0.056 &  0.615 $\pm$ 0.112 \\
NGC 7738 &   8.689 $\pm$ 0.031 &  -0.125 $\pm$ 0.041 &  -3.318 $\pm$ 0.110 &  0.316 $\pm$ 0.144 &  40.199 $\pm$ 0.524 & -1.097 $\pm$ 0.684 \\
NGC 7819 &   8.650 $\pm$ 0.009 &  -0.272 $\pm$ 0.012 &  -2.882 $\pm$ 0.031 & -0.112 $\pm$ 0.044 &  38.088 $\pm$ 0.087 &  1.484 $\pm$ 0.123 \\
IC 776  &    8.217 $\pm$ 0.008 &  -0.074 $\pm$ 0.015 &  -3.193 $\pm$ 0.026 & -0.116 $\pm$ 0.045 &  39.860 $\pm$ 0.167 & -0.179 $\pm$ 0.300 \\
IC 1256  &   8.707 $\pm$ 0.012 &  -0.316 $\pm$ 0.019 &  -2.871 $\pm$ 0.044 & -0.013 $\pm$ 0.068 &  37.536 $\pm$ 0.157 &  1.745 $\pm$ 0.246 \\
IC 5309  &   8.495 $\pm$ 0.030 &  -0.044 $\pm$ 0.043 &  -3.015 $\pm$ 0.066 &  0.220 $\pm$ 0.095 &  38.759 $\pm$ 0.223 & -0.355 $\pm$ 0.322 \\
UGC 8733 &   8.428 $\pm$ 0.005 &  -0.199 $\pm$ 0.009 &  -2.974 $\pm$ 0.017 & -0.172 $\pm$ 0.028 &  39.262 $\pm$ 0.052 &  0.896 $\pm$ 0.131 \\
UGC 12224 &  8.587 $\pm$ 0.032 &  -0.226 $\pm$ 0.054 &  -2.680 $\pm$ 0.103 & -0.256 $\pm$ 0.172 &  37.458 $\pm$ 0.256 &  2.164 $\pm$ 0.427 \\
UGC 12816 &  8.478 $\pm$ 0.010 &  -0.187 $\pm$ 0.016 &  -2.942 $\pm$ 0.026 & -0.057 $\pm$ 0.042 &  38.799 $\pm$ 0.067 &  0.845 $\pm$ 0.123 \\
\hline
\end{tabular}\\
\end{center}
\begin{flushleft}
\end{flushleft}
\end{table*}


\section{Results and Discussion}
\label{res-disc}

\begin{figure}
\begin{center}
\includegraphics[angle=0,width=0.48\textwidth]{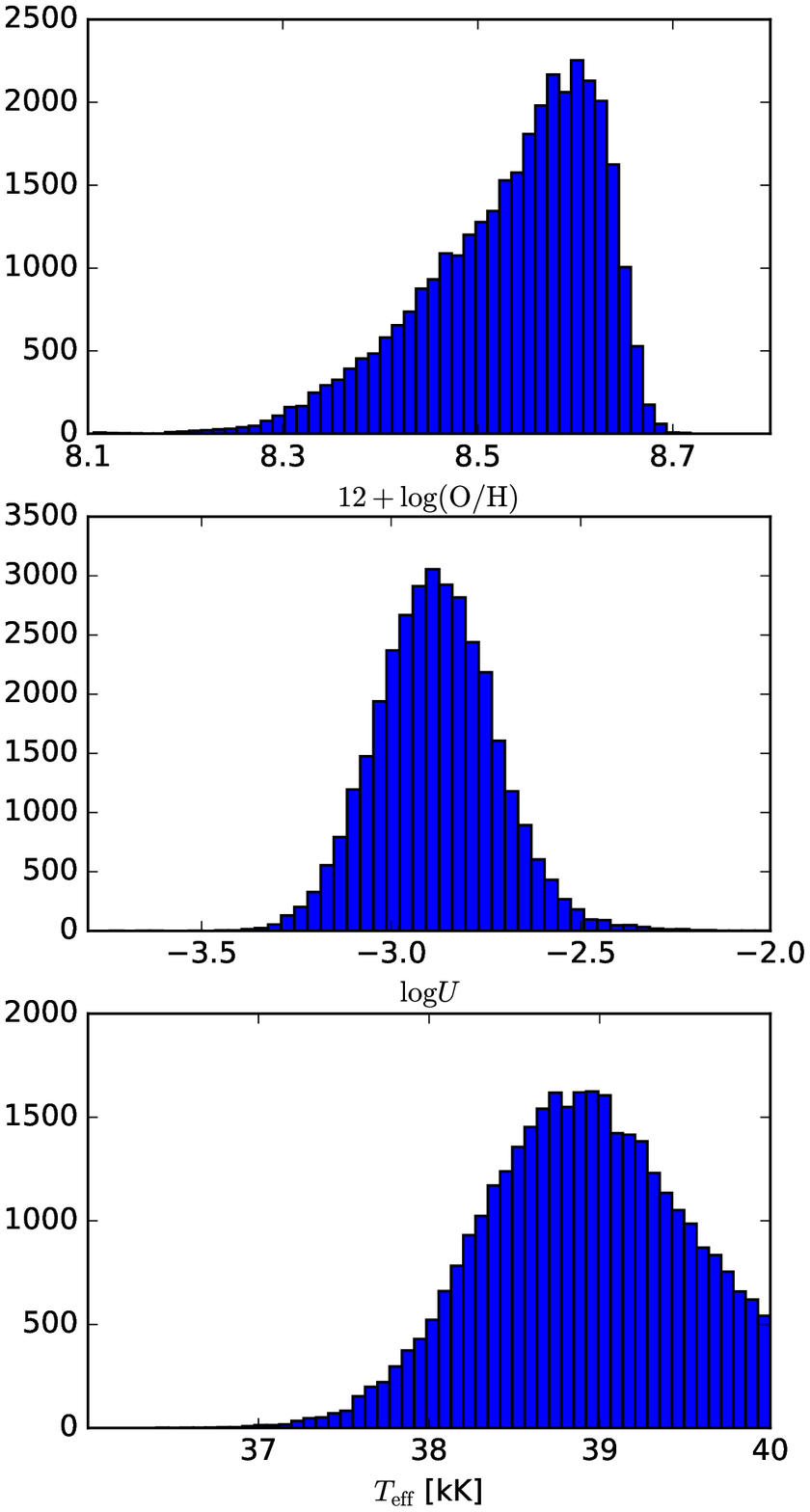}
\end{center}
\caption{ Histograms containing the oxygen abundances (left panel), logarithm of the ionization
parameter (middle panel) and effective temperature values (rigth panel) 
for the sample of objects presented in Section~\ref{data}.}
\label{fig1_oli}
\end{figure}


In Fig.~\ref{fig1_oli}, we present three histograms  containing the oxygen abundances,
logarithm of the ionization parameter and $T_{\rm eff}$ values obtained for the objects in our sample applying the methodology described above.
We can see that the oxygen abundance values (left panel) are 
in the range  $\rm 8.2 \: \la \: 12+\log(O/H) \: \la \: 8.8$
[equivalent to $ 0.3 \: \la \:  (Z/Z_{\odot}) \: \la \: 1.3$], with most objects  
presenting 12+log(O/H) around of 8.6 [$(Z/Z_{\odot})\approx0.8$]. 

Regarding the distribution of the ionization parameter (Fig.~\ref{fig1_oli}, middle panel),
it is in the $-3.4  \: \la \: \log U \: \la \: -2.5$ range, with an average value of about  $-2.8$.
Higher values for the ionization parameter than the ones derived by us seem to be most frequently found in objects
with low metallicity, such as the $U$ values ($\approx -1.3$ dex) derived by \cite{Lagos2018} for the central parts of 
the star-forming dwarf galaxies UM\,461 and Mrk\,600.

Concerning the effective temperature, it  can be seen in Fig.~\ref{fig1_oli}  (rigth panel) 
that most of the estimated $T_{\rm eff}$ values are around of  39 kK, 
with an average value of $38.5 \pm 1.0$ kK, being the scatter in the order of 
the uncertainty of our method, i.e. 2.5 kK (see Paper I). It should be noted
that this estimation for the uncertainty is an upper limit 
of the uncertainty for a single star-forming region, while in this work we use
many data points to estimate the distribution of $T_{\rm eff}$.
As described above, there is an artificial cut in $T_{\rm eff}$ at 40 kK due to the 
applied method (see Paper I). Nevertheless,  star-forming regions can be ionized 
by stars with $T_{\rm eff}$ higher than 40 kK. For example, \citet{Morisset2016} and \citet{Stasinska1996}
compared results of a grid of photonization models with observational data of star-forming regions. They estimated
the slopes of the SEDs of the ionizing sources, defined as the ratio between the number of neutral hydrogen ($\rm H^{0}$) and helium ($\rm He^{0}$) ionizing photons:
$Q_{o/1}=Q({\rm H^{0}})/Q(\rm He^{0})$ (a kind of softness parameter).  These estimated slopes are in the 0.1-1.0 range, which translates
into $T_{\rm eff}$ close to or above 40 kK. It is worth mention that only for few 
objects  the $T_{\rm eff}$ estimated values were higher than 40~kK, i.e.
most of the objects present $T_{\rm eff}$ values in the 30-40 kK range. This result is in agreement with recent $T_{\rm eff}$ estimations by \citet{RamirezAgudelo2017}, who used ground-based optical spectroscopy 
obtained in the framework of the VLT-FLAMES Tarantula Survey (VFTS) to determine parameters of 72 single O-type stars.

Estimations of  $T_{\rm eff}$  and consequently the $T_{\rm eff}$ gradients  are dependent on the stellar atmosphere model assumed in the 
photoionization models (e.g.\ \citealt{Stasinska1997, Dors2003, Morisset2004b, Morisset2004}) and on the match between the metallicity of the 
atmosphere models and the gas \citep{Morisset2004}. In our case, the $T_{\rm eff}$-$R$  relation was derived assuming
the WM-basic stellar atmosphere models \citep{Pauldrach2001}, which are available only for two metallicities:
solar and half solar. Therefore, there is an inconsistent match between the stellar and gas metallicities for
the photoionization models with $(Z/Z_{\odot})$=0.03 and 0.2, which does overpredict the $T_{\rm eff}$ value for low metallicity objects,
reducing the  effective gradient generally found in spiral galaxies \citep{Morisset2004, Dors2011}. However, as can be seen
in Fig.~\ref{fig1_oli} (middle panel), most of the objects ($\approx82\%$) present
$\rm 12+log(O/H) \: \ga \:8.4$ $[\mathrm{i.e.\ } (Z/Z_{\odot})  \: \ga \: 0.5]$. Therefore, the missmatch between the stellar and gas metallicities in 
low metallicity models has little effect on our $T_{\rm eff}$ estimations.

\begin{figure}
\begin{center}
\includegraphics[angle=0,width=0.8\columnwidth]{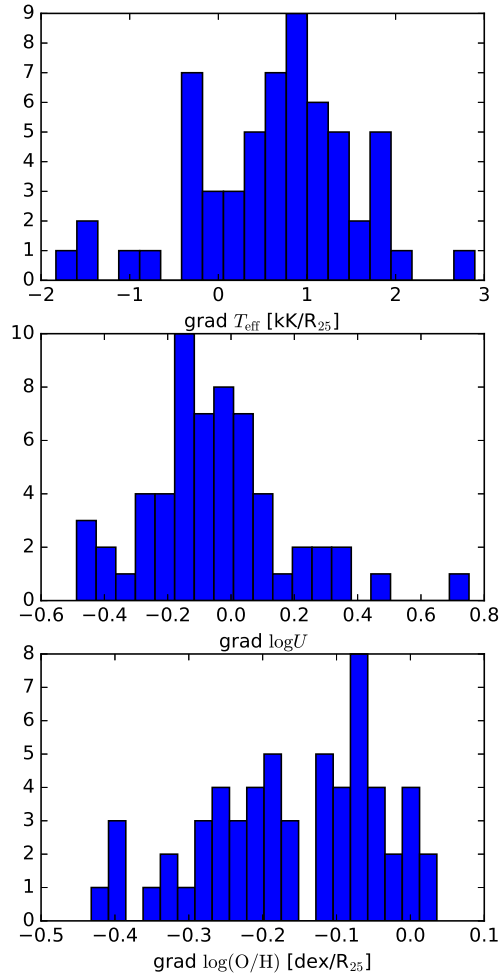}
\end{center}
\caption{Histograms containing estimated values for the gradients of: $T_{\rm eff}$ (upper panel),
$\log U$ middle panel),
and $\log(\rm O/H)$ (lower panel)
for our sample of galaxies.}
\label{Teff_U}
\end{figure}

In Fig.~\ref{Teff_U} we present histograms containing the estimated values for gradients of: $T_{\rm eff}$, $\log U$, 
and oxygen abundance for 
our sample of galaxies. We can see that for most of the galaxies ($\sim70$ \%) positive
values of $T_{\rm eff}$ gradient are derived, despite of a number of flat and negative gradients.
The median value of the $T_{\rm eff}$ radial gradients is 0.762 kK/R$_{25}$, the minimum and 
the maximum values are -1.8 and 2.9, respectively.
The prevalence of positive $T_{\rm eff}$ gradients is compatible with what is expected under 
the hypothesis that stars are formed following an Initial Mass Function (IMF) with an universal upper mass limit ($M_{\rm up}$) and 
the variation of the $T_{\rm eff}$ with the galactocentric distance is due to 
line blanketing effects taking place in the stellar atmospheres. Alternatively, 
this prevalence could be due to an increment in the $M_{\rm up}$ of the IMF (and then its $T_{\rm eff}$) as the metallicity decreases.

\citet{Dors2017} analyzed  $T_{\rm eff}$ in a small sample of 14 spiral galaxies and also found 
that most of the galaxies ($\sim80$ \%) presents positive gradients while others show flat ($\sim15$ \%) 
or negative ($\sim5$ \%) slopes. Similar results were found by \citet{PerezMontero2009}, who studied the behaviour
of the $\eta'$ parameter (sensitive to $T_{\rm eff}$) along the disk of 12 galaxies. 
Therefore, in consonance with \citet{Dors2017} and \citet{PerezMontero2009}, we found that although 
positive  $T_{\rm eff}$ gradients are present in the disk of most of spiral galaxies, 
this is not an universal property. 

In the middle panel of Fig.~\ref{Teff_U} we can note that both negative 
and positive gradients of $\log U$ are derived for our sample of galaxies, with a tendency to derive negative gradients more frequently presenting a median value of -0.1~dex/R$_{25}$ and being in the range from -0.5 to 0.8~~dex/R$_{25}$. 
The overwhelming majority of galaxies in our sample have negative oxygen abundance gradients being in the range from -0.43 to 0.04~dex/R$_{25}$ (see lower panel of the same figure). The median value of the oxygen abundance gradient is -0.15~dex/R$_{25}$.
 
\begin{figure}
\begin{center}
\includegraphics[width=1.0\linewidth]{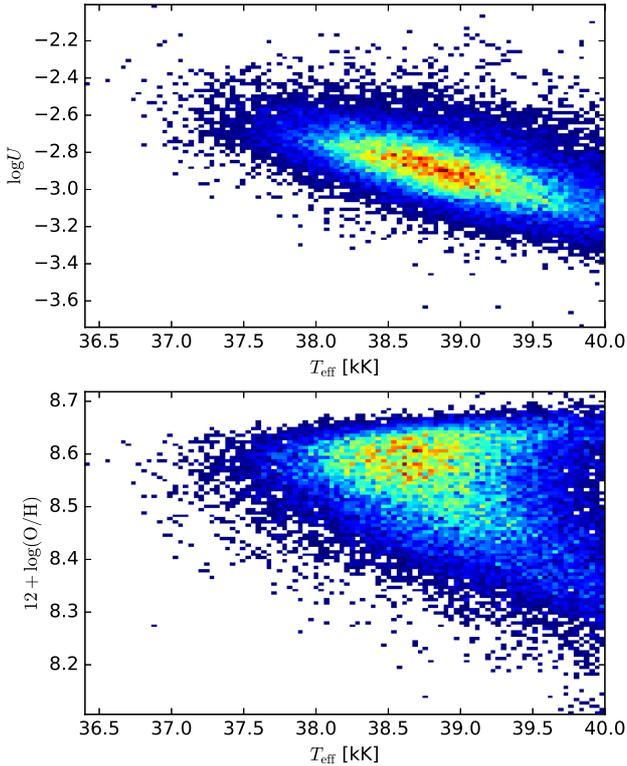}
\end{center}
\caption{Density maps for the individual spaxels of our sample of CALIFA galaxies.
{\it Top panel.} $\log U$ as a function of $T_{\rm eff}$.
{\it Bottom panel.} Oxygen abundance as a function of $T_{\rm eff}$.}
\label{correl}
\end{figure}

To investigate the correlation between $T_{\rm eff}$  and the other studied nebular parameters: 
$\log U$ and 12+$\log(\rm O/H)$, we plot in Fig.~\ref{correl} 
these parameters as a function of $T_{\rm eff}$ for the individual spaxels of  our sample. 
Top panel of this figure shows that $\log U$ decreases as $T_{\rm eff}$ increases. 
This result is in agreement with the one derived by \citet{Morisset2016}, who found that 
$Q_{0/1}$ (wich is inversely proportional to $T_{\rm eff}$)  is increasing with $\log U$.
This result indicates that cooler stars lead to higher $U$,  in contradiction with the assumption 
that  H\,{\sc ii} regions ionized by hotter stars would have higher $U$  because these stars are emitting more ionizing photons.
\citet{Sanders+16}  showed that the ionization parameter has a weak dependence on both
the rate of ionizing photon production and the gas density, and is somewhat more sensitive to the volume filling factor ($\epsilon$):
\begin{equation}
U \: \propto Q({\rm H})^{1/3} \:  N_{\rm e}^{1/3} \: \epsilon^{2/3}.
\end{equation}
Therefore, it is possibly that nebulae ionized by 
cooler stars present higher $\epsilon$ (and consequentely higher $U$) than those ionized by hotter stars.

Despite of the conclusions by \citet{Shields1978,Vilchez1988,Henry1995,Dors2003,Dors2005}, 
who claimed that high metallicity H\,{\sc ii} regions have lower $T_{\rm eff}$ compared 
to those with low metallicity,  we do not  find any clear  correlation between $T_{\rm eff}$ 
and 12+$\log(\rm O/H)$ for the spaxels of all galaxies in our sample (see also \citealt{Morisset2004, Dors2017}). 
Contradiction between previous and current results can be caused by the fact that the $T_{\rm eff}$ -- metallicity
relation is not unique, i.e. this relation is different for 
the different galaxies. This suggestion will be discussed below.

\begin{figure}
\begin{center}
\includegraphics[width=0.99\linewidth]{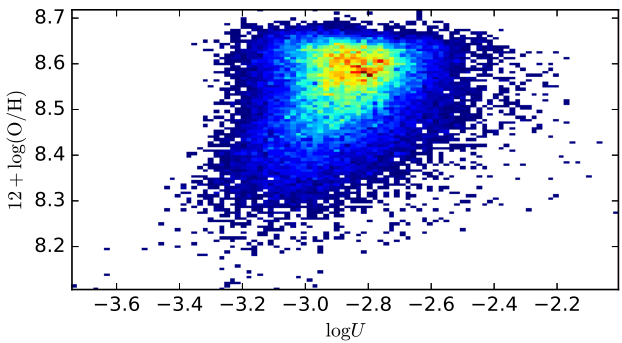}
\end{center}
\caption{Same as Fig.~\ref{correl} for 12+$\log(\rm O/H)$ as a function of $\log U$.}
\label{correl2}
\end{figure}

In Fig.~\ref{correl2} we plotted 
12+$\log(\rm O/H)$ versus $\log U$ where a large scatter is noted and not apparent correlation can be seen between both parameters.
This result is in consonance with, for example, the one found by \citet{Dors2011}, who derived  oxygen abundances and ionization parameters  
from  diagnostic diagrams containing photoionization model results and observational data of  H\,{\sc ii} regions.
\citet{Kaplan2016} presented a study of the excitation conditions and metallicities in eight
nearby spiral galaxies from the VIRUS-P Exploration of Nearby Galaxies (VENGA) survey. These authors
calculated the ionization parameter by using  an  iterative determination proposed by \citet{Kewley2008}, and they
did not notice any clear trends between  $U$ and $Z$  (see also \citealt{LaraLopez2013}). In other hand, a trend  for  H\,{\sc ii} regions showing that those with
higher values of $\log U$ present lower metallicities was derived, for example, by \citet{Morisset2016}, 
who used a large grid of photoionzation models in  order to reproduce emission line intensities also taken from the CALIFA database
(see also \citealt{Enrique2017} and references therein).   The  relation between ionization paremeter and oxygen abundance seems  to be dependent on the methodology  employed to calculate these parameters 
(e.g. \citealt{Kruhler2017}) or on the geometry assumed in the photoionization models 
(see, for example, Fig.~13 of \citealt{Morisset2016}.)

\begin{figure}
\begin{center}
\includegraphics[width=1.0\linewidth]{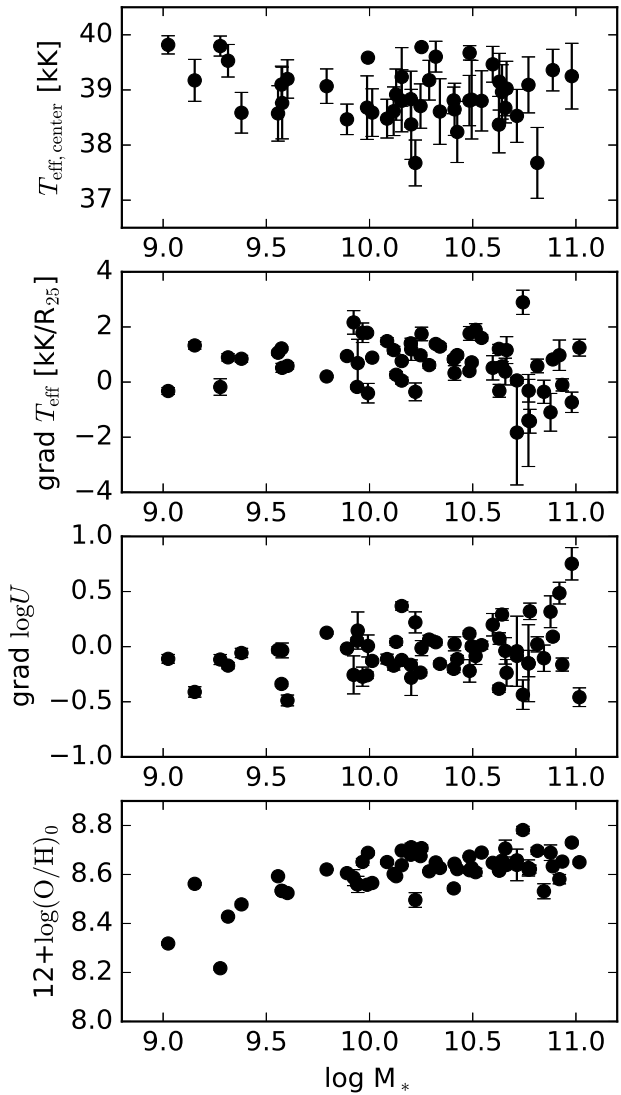}
\end{center}
\caption{Panels from top to bottom: central $T_{\rm eff}$ (average values), 
radial gradients of $T_{\rm eff}$, radial gradients of $\log U$, and 12+$\log(O/H)_0$ as 
a function of the stellar mass of the galaxy.
}
\label{figure:hist}
\end{figure}

Fig.~\ref{figure:hist} shows the central $T_{\rm eff}$,  gradients of $T_{\rm eff}$ and $\log U$, 
and 12+$\log(\rm O/H)_0$ as a function of the stellar mass of the galaxies. Central $T_{\rm eff}$ 
was calculated as the average values for the spaxels with $R < 0.2 R_{25}$. We found no correlation 
between central $T_{\rm eff}$, $T_{\rm eff}$ and $\log U$ gradients and the stellar mass of the galaxies.

\begin{figure*}
\begin{center}
\includegraphics[width=0.9\linewidth]{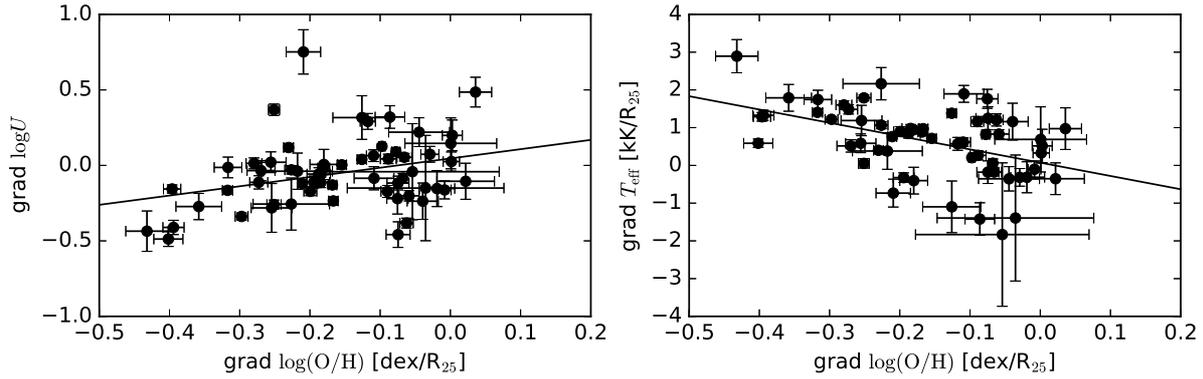}
\end{center}
\caption{{\it Left panel:} Radial gradient of $\log U$ as a function of the oxygen abundance radial gradient.
{\it Right panel:} Radial gradient of $T_{\rm eff}$ as a function of the
oxygen abundance radial gradient. Solid lines are the best fit linear regression to the data.}
\label{figure:grad}
\end{figure*}

The radial gradients of $\log U$ and  $T_{\rm eff}$ as a function of oxygen abundance 
gradient are shown in Fig.~\ref{figure:grad}, finding a correlation in both cases. The former presents a positive correlation, the gradient of $\log U$ increases 
as the $\log(\rm O/H)$ gradient increases. As the p-value is 0.0001, null hypothesis that there is no correlation between $\log U$ and ${\rm O/H}$ should be rejected.
On the other hand, an anti-correlation between the gradient of $T_{\rm eff}$ and the oxygen abundance gradient is clearly seen for our sample with a p-value of $10^{-5}$. 
Moreover, galaxies with flat oxygen abundance gradients tend to have flat $\log U$ and $T_{\rm eff}$ gradients too.
Therefore, one can expect an anti-correlation between the $T_{\rm eff}$ and the oxygen abundance.
Indeed, such anti-correlation can be seen in the low metallicity zone on the bottom panel of Fig.~\ref{correl}. However, at the high metallicity regime there is no correlation between $T_{\rm eff}$ and $\log(\rm O/H)$ for individual spaxels. 
Using the softness parameter defined by \cite{Vilchez1988} as 

\[
\eta'\,=\,\frac{([\textrm{O\,{\sc ii}}]\lambda\lambda 3727,3729/[\textrm{O\,{\sc  iii}}]\lambda\lambda 4959,5007)}{([\textrm{S\,{\sc ii}}]\lambda\lambda 6717,6731/[\textrm{S\,{\sc iii}}]\lambda\lambda 9069,9532)},
\]
which works as a diagnostic for the nature and the efective temperature of the ionizing radiation 
field \cite[see][]{Diaz+85}, \cite{Diaz+07} and \cite{Hagele08} compared the ionization structure 
for star-forming regions in different environments. They found that high metallicity Circumnuclear 
Star Forming Regions (CNSFRs) segregates from high metallicity  disk H\,{\sc ii} regions, with 
the former showing $T_{\rm eff}$ values of about 40 kK and the last ones of about 35 kK, 
a temperature range similar to those found by us for the objects in the high metallicity regime. 
On the other hand, these authors also found that the low metallicity H\,{\sc ii} galaxies belonging 
to their sample \cite[see also the  H\,{\sc ii} galaxies studies by][]{Hagele+06,Hagele+08,Hagele+11,Hagele+12,PerezMontero+10} 
present a similar behaviour and $T_{\rm eff}$ values (40 kK) than those shown by the CNSFRs. 
In all the cases the low metallicity H\,{\sc ii} galaxies show high $T_{\rm eff}$ values, 
in agreement with the results derived from Fig.\ \ref{correl}.
A possible explanation of this fact could be that, for an individual galaxy, the $T_{\rm eff}$ increases as the oxygen abundance decreases 
but there is not a unique $T_{\rm eff}$ -- $\log(\rm O/H)$ relation for all galaxies.

\begin{figure}
\begin{center}
\includegraphics[width=1.0\linewidth]{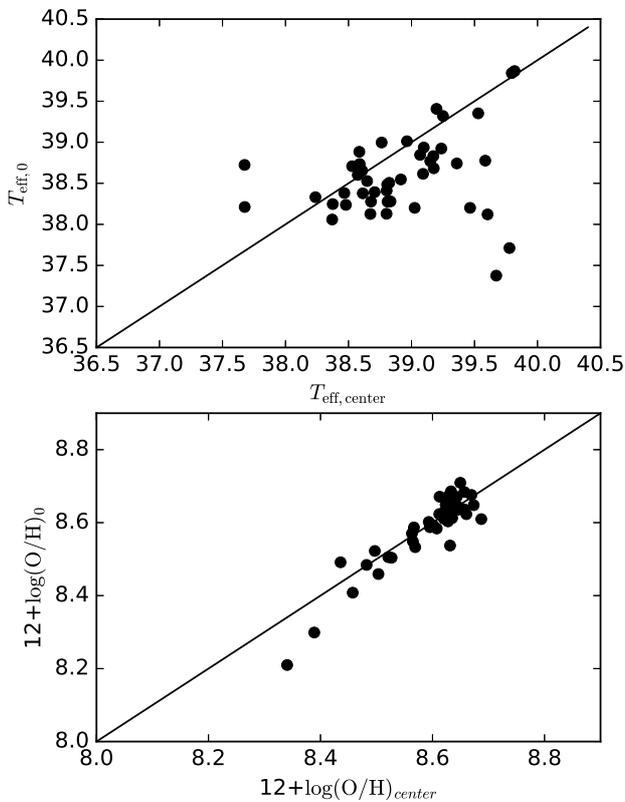}
\end{center}
\caption{{\it Upper panel:} Comparison of $T_{\rm eff, 0}$  and $T_{\rm eff, center}$.
{\it Bottom  panel:} Relation between 12+log(O/H)$_{0}$ and 12+log(O/H)$_{center}$. 
In both panels the one-to-one relation is shown.}
\label{figure:Teffcomp}
\end{figure}

Finally, the averaged $T_{\rm eff}$ for the spaxels with $R \: < \: 0.2R_{25}$ ($T_{\rm eff, 0}$) as a function of the $T_{\rm eff}$ value extrapolated to $R=0.1R_{25}$ ($T_{\rm eff, center}$) is plotted in the upper panel of Fig.~\ref{figure:Teffcomp}.
It shows that for the galaxies in our sample the $T_{\rm eff, 0}$ 
can be significantly lower than the $T_{\rm eff, center}$.
This fact could be considered as an indication that the star formation  processes at the central parts of galaxies, which determine the $T_{\rm eff}$, are not similar to those along the disks. This could be due to metallicity effects, in the sense that metallicities at the center could be higher than the ones expected extrapolating the radial O/H gradient, which leads to the cooling of the atmospheres of massive stars.
However, the behaviour of the $\rm (O/H)_{0}$ versus $\rm (O/H)_{center}$ for the galaxies in our sample 
(bottom panel of Fig.~\ref{figure:Teffcomp}) does not show significant bias between $\rm (O/H)_{0}$ and 
$\rm (O/H)_{center}$. Thus, another effect rather than metallicity seems to be responsible for producing the discrepancy between  $T_{\rm eff, 0}$ and $T_{\rm eff, center}$, since not only the metallicity controls the $T_{\rm eff}$. 
Star formation processes in nuclear regions of galaxies can be 
altered by, for example, supernova explosions and/or the presence of Wolf Rayet stars  (most common in high metallicity environments).
Moreover, the  gas outflows found in nuclear starbursts and in Active Galaxy Nuclei, that extends 
on kiloparsec scales, could potentially suppress star formation in their host galaxies \citep{Gallagher2018}
modifying the $T_{\rm eff}$ expected from the radial gradient and producing the discrepancy seen in Fig.~\ref{figure:Teffcomp}. 
Also, gas flux from outskirts parts of the disks
(e.g. \citealt{Rosa2014}) could be falling into the nucleus modifying the star formation processes.

\section{Conclusion}
\label{conc}

We used homogeneous spectroscopic data of H\,{\sc ii} regions taken from the CALIFA survey 
and a theoretical calibration between the effective temperature of ionizing star(s)
($T_{\rm eff}$) and the ratio $R$ = $\log$([O\,{\sc ii}]$\lambda\lambda$3727,3729/[O\,{\sc iii}]$\lambda$5007)
to investigate the universality of $T_{\rm eff}$ gradients in spiral galaxies as well as
correlation between $T_{\rm eff}$, the ionization parameter ($U$) and the oxygen abundance of H\,{\sc ii} regions.
We found that most of the galaxies in our sample ($\sim 70$ \%) presents positive $T_{\rm eff}$ radial gradients, with a median value of 0.762 kK/$R_{25}$, even though some galaxies exhibit negative or flat $T_{\rm eff}$ radial gradients. Therefore, we conclude that $T_{\rm eff}$ gradients
are not an universal property of spiral galaxies. We also found that radial gradients 
of both $\log U$ and $T_{\rm eff}$ depend on the oxygen abundance gradient, 
in the sense that the gradient of $\log U$ increases as the $\log({\rm O/H})$ gradient increases while the $T_{\rm eff}$ gradient decreases as the $\log({rm O/H})$ increases. 
Moreover, galaxies with flat oxygen abundance gradient tend to have flat $\log U$ and $T_{\rm eff}$ gradients.

\section*{Acknowledgements}

We are grateful to the referee for his/her constructive comments. \\
I.A.Z.\ thank  FAPESP for the financial support during his visit
to UNIVAP (FAPESP grant number 2017/19538-1). \\
OLD and ACK thank FAPESP and CNPq. 
I.A.Z. acknowledges the support by the Ukrainian National Grid 
project (especially project 400Kt) of the NAS of Ukraine. \\
This study uses data provided by the Calar Alto Legacy Integral Field 
Area (CALIFA) survey (http://califa.caha.es/).
Based on observations collected at the Centro Astronomico Hispano Aleman 
(CAHA) at Calar Alto, operated jointly by the Max-Planck-Institut fur 
Astronomie and the Instituto de Astrofisica de Andalucia (CSIC).

\bibliography{reference}

\end{document}